\documentclass[
superscriptaddress,
showkeys,
showpacs,
amsmath,amssymb,
prl,
twocolumn,
]{revtex4-2}

\usepackage{lineno}
\usepackage[colorlinks,linkcolor=red,citecolor=blue,urlcolor=blue]{hyperref}
\usepackage{amsfonts}
\usepackage{amssymb}
\usepackage{graphicx}
\usepackage{epstopdf}
\usepackage{pgfplots}
\pgfplotsset{compat=1.14}
\usepackage{bm}
\usepackage{soul}
\usepackage{comment}

\usepackage[draft]{todonotes}

\begin{document}
	\title{Asymptotic freedom in the lattice Boltzmann theory}
	\author{S. A. Hosseini}
	\affiliation{Department of Mechanical and Process Engineering, ETH Zurich, 8092 Zurich, Switzerland}%
	\author{I. V. Karlin}\thanks{Corresponding author}
 \email{ikarlin@ethz.ch}
	\affiliation{Department of Mechanical and Process Engineering, ETH Zurich, 8092 Zurich, Switzerland}%
	\date{\today}
	\begin{abstract}
		{
			Asymptotic freedom is a feature of quantum chromodynamics  that guarantees its well-posedeness. We derive an analog of asymptotic freedom  enabling unconditional stability of lattice Boltzmann simulation of hydrodynamics. 
			For the lattice Boltzmann models of nearly-incompressible flow, we show that
			the equilibrium based on entropy maximization is uniquely renormalizable. 
			This results in a practical algorithm of constructing unconditionally stable lattice Boltzmann models.
   }
	\end{abstract}
	\maketitle

{The lattice Boltzmann method (LBM) became a popular tool for the simulation of complex fluid dynamics, with applications ranging from turbulent flows to multiphase and multicomponent flows, combustion and relativistic flows \cite{succi_lattice_2002,kruger_lattice_2017}. 
In LBM, a simple kinetic equation of Boltzmann type for the populations of a controlled number of designer particles’ velocities, forming links of a regular spatial lattice, is solved numerically in a "stream-along-links and relax-to-equilibrium at the nodes" fashion. 
Efficiency and universality are keywords one associates with LBM.  
At the same time, theoretical foundations of LBM remain obscure in lieu of  long-standing issues of stability and accuracy.
}

{In this Letter, 
we propose a new approach to LBM by following ideas of renormalization group \cite{peskin2018introduction,zinn2021quantum,kunihiro_2022}.
We derive  {necessary} and {sufficient} conditions of linear stability and identify coupling parameters governing these conditions.  
We find that unconditional stability implies vanishing coupling at large flow velocity, which bears direct analogy to asymptotic freedom in quantum chromodynamics \cite{gross1973ultraviolet,politzer1973reliable,gross1973asymptotically}.
We show that this change of paradigm in deriving equilibria leads to unconditionally stable lattice Boltzmann models.}

We consider the  lattice Bhatnagar--Gross--Krook (LBGK) model \cite{qian1992lattice} for nearly-incompressible flows,
\begin{equation}
	f_i(\bm{r}+\bm{c}_i \delta t, t+\delta t) - f_i(\bm{r}, t)= 2\beta\left( f_i^{\rm eq}(\rho, \bm{u}) - f_i\right).\label{eq:LBGK}
\end{equation}
Here $f_i$ are the populations of the discrete velocities $\bm{c}_i$, $i=1,\dots, Q$, $\bm{r}$ is the position in space, $t$ is the time, $\delta t$ is the time-step, $\rho$ is the fluid density and $\bm{u}$ is the flow velocity,
\begin{align}
	&\sum_{i=1}^Q \{1,\bm{c}_i\}f_i = \{\rho,\rho\bm{u}\},
\end{align}
Furthermore, $\beta\in[0,1]$ is the relaxation parameter
which is tied to the viscosity, 
\begin{align}
	\nu=\varsigma^2\delta t\left(\frac{1}{2\beta}-\frac{1}{2}\right),\label{eq:viscosity}
\end{align}
where $\varsigma=c_s\delta r/\delta t$ is the lattice speed of sound, $\delta r$ is the lattice spacing while $c_s$ is a pure constant dependent on the choice of the lattice. Below, we use lattice units by setting $\delta r=\delta t=1$ and 
consider the standard first-neighbor lattices in space dimension $D$ defined as a $D$-fold tensor product of one-dimensional velocities $c_{i\alpha}\in\{-1,0,1\}$. These are the D$D$Q$3^D$ lattices characterized by the lattice speed of sound, 
\begin{equation}
		\varsigma = \frac{1}{\sqrt{3}}.\label{eq:cs_lattice}
\end{equation}
A generic class of equilibria $f_i^{\rm eq}$ is the focus of our study: First, we introduce a triplet of functions $\Psi_{i\alpha}(\xi,\mathcal{P})$, $i\alpha\in\{-1,0,1\}$: 
	$\Psi_0=1-\mathcal{P}$, $\Psi_{-1}=(1/2)(-\xi+\mathcal{P})$, $\Psi_{1}=(1/2)(\xi+\mathcal{P})$. For a $D$-dimensional lattice, equilibria are defined by a product form,
    \begin{equation}\label{eq:generic_isothermal_EDF}
	f_i^{\rm eq} (\rho,\bm{u})=  \rho  \prod_{\alpha=1}^{D} \Psi_{i\alpha}(u_\alpha,\mathcal{P}^{\rm eq}_{\alpha\alpha}).
\end{equation}
Here $\mathcal{P}^{\rm eq}_{\alpha\alpha}$ are diagonal component of the equilibrium pressure tensor at unit density, 
\begin{align}
	\mathcal{P}^{\rm eq}_{\alpha\alpha}= \pi^*_{\alpha\alpha} + u_\alpha^2.\label{eq:Peq_gen}
\end{align}
{LBGK setup becomes complete once the function $\pi^*_{\alpha\alpha}$ is specified. At this point, a common LBM  \cite{qian1992lattice,succi_lattice_2002,kruger_lattice_2017} suggests {isotropic} pressure,
	$\pi_{\alpha\alpha}^{*}=\varsigma^2$.
While this  
seems natural and is  equivalent to \eqref{eq:generic_isothermal_EDF} matching moments of the classical Maxwellian, {it should be reminded that Galilean invariance of LBM  is restricted to small flow velocities,} as implied by the lattice constraint, $c_{i\alpha}^3=c_{i\alpha}$.}
{Hence, in order to derive the pressure in a rigorous fashion, we follow a path inspired by renormalization group and define a "space of theories" by assuming that the function $\pi^*_{\alpha\alpha}\ge 0$ {may} depend on the flow velocity component $u_{\alpha}$.
A coarse-graining will be performed (the Chapman--Enskog calculation \cite{chapman1990mathematical}) to identify the necessary stability conditions at fixed point (hydrodynamic limit) and their implication for the pressure $\pi^*_{\alpha\alpha}$.}

\begin{figure}[t!]
	\centering
	\includegraphics[width=0.8\linewidth,keepaspectratio]{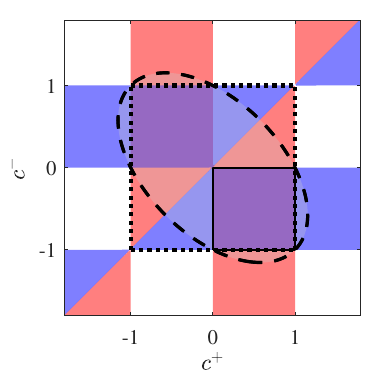}
	\caption{Positivity domain of attenuation rates  $\mathcal{R}^{\pm}$ \eqref{eq:attenuation_rates_1D} and of the viscosity factor $\mathcal{A}$ \eqref{eq:A} as a function of eigen-modes $c^+$ and $c^-$. Red: Positivity domain  of  $\mathcal{R}^{+}$; Blue: Positivity domain of $\mathcal{R}^{-}$. Purple: Positivity domain of both $\mathcal{R}^{+}$ and $\mathcal{R}^{-}$ simultaneously. Positive square root convention in Eq.\ \eqref{eq:modes_D1} restricts stability domain to the bottom-right quadrant \eqref{eq:condition1}, shown with solid black lines. Area inside dashed black elliptic contour: Positivity domain of the viscosity factor $\mathcal{A}$ \eqref{eq:A}. Area inside black dotted square: Validity domain of CFL condtion.}
	\label{Fig:attenuation_positivity_csp_csm}
\end{figure}

{We begin with the one-dimensional LBGK on the D$1$Q$3$ lattice.
The first Chapman--Enskog approximation  results in the continuity and momentum equations,  
  $\partial^{(1)}_t\rho+\partial_x \rho u=0$, 
$\partial^{(1)}_t \rho u+\partial_x \rho u^2+\partial_x\rho\pi^*=0$, where we omitted index $x$ to ease notation.  In order to identify the coupling parameters, we perform the characteristics analysis \cite{Hosseini_Karlin_PRE_2023} which reveals a pair of normal {eigen-modes} propagating with the speeds $c^{\pm}$,}
	\begin{align}
		c^{\pm}= u + \frac{1}{2}\partial_u\pi^*
		\pm \sqrt{\frac{1}{4}\left(\partial_u\pi^*\right)^2+  \pi^{*}},\label{eq:modes_D1}
	\end{align}
where positive square root is assumed, $c^+-c^->0$.
Conversely, upon defining the two sound speeds, $\varsigma^{\pm}=c^{\pm}-u$, the pressure and its derivative are expressed as,
\begin{align}
	&\pi^*=-\varsigma^+\varsigma^-,\label{eq:pistarvarsigma}\\
	&\partial_u\pi^*=\varsigma^+ +\varsigma^-.\label{eq:dupistarvarsigma}
\end{align}
In the second Chapman--Enskog approximation, the momentum equation is modified by a nonequilibrium term,
    $\partial_t^{(2)}\rho u =- \partial_x{\pi}^{\rm neq}$,
with the nonequilibrium diagonal component of the pressure tensor as,
\begin{equation}\label{eq:noneq_second_order_moment}
    {\pi}^{\rm neq} = -\left(\frac{1-\beta}{2\beta}\right) \left(2 \mathcal{A} \rho \varsigma^2 \partial_x u + \mathcal{B} \partial_x \rho\right).
\end{equation}
Here $\mathcal{A}$ ({viscosity factor}) and $\mathcal{B}$ ({compressibility error}) are expressed in terms of the pressure and its derivative,
\begin{align}
		&\mathcal{A}=\frac{1}{2\varsigma^2}\left(3\varsigma^2-3u^2-\pi^* -\partial_u\pi^*\left(3u+\partial_u\pi^*\right)\right), 
		\label{eq:A}\\
			&\mathcal{B}=-\left(3u + \partial_u\pi^* \right)\pi^*+3u\varsigma^2-u^3.
		\label{eq:B}
	\end{align}
{Finally, we perform the spectral analysis of the hydrodynamic equations, linearized around $\{\rho,u\}$, to find the following leading-order {dissipation-dispersion relations} for the eigen-frequencies,
    \begin{align}
    	\omega^{\pm} &= c^\pm k + \mathrm{i} \nu \mathcal{R}^{\pm} k^2 + O\left(k^3\right),
    	\label{eq:omega1D}
    \end{align}
where $k$ is the wave vector, $\mathrm{i}=\sqrt{-1}$, and $\nu$ is the viscosity \eqref{eq:viscosity}, while $\mathcal{R}^{\pm}$ are attenuation rates, written in terms of eigen-modes,
\begin{equation}
	\mathcal{R}^{\pm}=\pm \frac{c^{\pm}\left(3 \varsigma^2 - (c^{\pm})^2\right)}{\varsigma^2(c^+-c^-)}.\label{eq:attenuation_rates_1D}
\end{equation}
Thus, spectral analysis of hydrodynamics-level equations reveals that both the dispersion and the dissipation are governed by two coupling parameters, either $\{\pi^*, \partial_u \pi^*\}$ or, equivalently and somewhat more symmetrically, $\{c^+,c^-\}$ or $\{\varsigma^+,\varsigma^-\}$. Note that, for the isotropic pressure, the derivative $\partial_u\pi^*$ vanishes, and the two-parametric coupling degenerates to a constant, $\varsigma^\pm=\pm\varsigma$. In a  general case however, the coupling parameters may depend on the flow velocity.}
{Hence, with the positivity of the viscosity $\nu$ already established by the bound on the relaxation parameter $\beta\in [0,1]$, the \emph{necessary} stability condition of the LBGK system in the long-wave limit $k\rightarrow 0$ is the positivity of attenuation rates \eqref{eq:attenuation_rates_1D} $\mathcal{R}^{\pm}\ge0$, see Fig.\ \ref{Fig:attenuation_positivity_csp_csm}. 
Positivity domain of the viscosity factor \eqref{eq:A}, written in terms of eigen-modes,  $\mathcal{A}=({1}/{2\varsigma^2})\left(3\varsigma^2-(c^+)^2-(c^-)^2 - c^+ c^- \right)$, is also shown in Fig.\ \ref{Fig:attenuation_positivity_csp_csm}.}
{With the lattice speed of sound \eqref{eq:cs_lattice}, we find that the attenuation rates \eqref{eq:attenuation_rates_1D} are non-negative if the eigen-modes \eqref{eq:modes_D1} satisfy the following inequalities:}
    \begin{align}
    0  \leq  c^+ \leq 1,\
     -1\leq  c^- \leq 0. \label{eq:condition1}
    \end{align}
{Note that the necessary stability condition \eqref{eq:condition1} is also consistent with (and is stronger than) the Courant--Friedrichs--Levy (CFL) condition \cite{courant_uber_1928}, 
    $\max\{\lvert c^\pm\rvert\}\leq 1$, which tells that no eigen-mode can propagate faster than the maximal speed equal to the lattice link, see Fig.\ \ref{Fig:attenuation_positivity_csp_csm}.
Moreover, with the explicit form of the eigen-modes \eqref{eq:modes_D1}, the stability condition \eqref{eq:condition1} translates into the following limits of the pressure and its derivative at the maximal flow velocity $\lvert u\rvert= 1$:
\begin{align}
&	\lim_{u\to \mp 1}\pi^*=0, \label{eq:af1}\\
&	1\leq \lim_{ u \to \mp 1}\left(\pm\partial_u\pi^*\right)\leq 2.\label{eq:af2}
\end{align}
In other words, the necessary stability condition \eqref{eq:condition1} for the slow modes of the D$1$Q3 lattice Boltzmann model implies vanishing pressure at the maximal flow velocity $\lvert u\rvert =1$. 
This can be interpreted as a case for {asymptotic freedom}: The necessary condition for linear stability of the lattice Boltzmann system is the asymptotic vanishing of the pressure $\pi^*$ in the limit of large fluid velocity. Isotropic pressure $\pi^*=\varsigma^2$ is not asymptotically free and violates the necessary stability condition \eqref{eq:condition1} at $|u^{\max}|=1-1/\sqrt{3}\approx 0.42$.} 
{Differently put, since Galilean invariance of LBM is limited to small flow velocities, the pressure "needs to bend" at velocities sufficiently far from $u=0$ and adjust in such a way as to maintain unconditional stability of the hydrodynamic limit \eqref{eq:condition1}.}

In order to find the asymptotically free pressure, we note 
a distinguished case when the compressibility error cancels in the nonequilibrium flux of momentum \eqref{eq:noneq_second_order_moment},
\begin{equation}
	\mathcal{B}=0. \label{eq:no_compressibility_error}
\end{equation}
With \eqref{eq:no_compressibility_error}, the D$1$Q$3$ LBGK model becomes {renormalizable}: Since $\mathcal{A}\ge0$ for the asymptotically free pressure, see Fig.\ \ref{Fig:attenuation_positivity_csp_csm}, 
the viscosity factor can be absorbed into the viscosity by renormalizing the relaxation parameter,
\begin{equation}
	\beta^*=\frac{\varsigma^2\mathcal{A}}{2\nu +\varsigma^2\mathcal{A}},\label{eq:renorm_beta}
\end{equation}
whereby Eq.\ \eqref{eq:noneq_second_order_moment} assumes a purely Navier--Stokes form,
	\begin{equation}\label{eq:renorm_noneq_second_order_moment}
		{\pi}^{\rm neq} = -\frac{1-\beta^*}{2\beta^*}  \rho\varsigma^2  (2\partial_x u).
	\end{equation}
With \eqref{eq:B}, the {no-compressibility-error} condition \eqref{eq:no_compressibility_error}  is a first-order ordinary differential equation, which admits unique solution subject to initial condition $\pi^*(0)=1/3$,
\begin{align}
	\pi^{*}& = \varsigma^2 \left( 2\sqrt{{1+(u/\varsigma)}^2 } -{1-(u/\varsigma)}^2\right).\label{eq:Pstar_entropic}
\end{align}
Direct evaluation verifies that \eqref{eq:Pstar_entropic} validates the inequalities \eqref{eq:condition1} and is thus asymptotically free.
Moreover, it is striking that, with the pressure \eqref{eq:Pstar_entropic}, the equilibrium \eqref{eq:generic_isothermal_EDF} coincides with the {entropic equilibrium} \cite{ansumali_stabilization_2000,ansumali_minimal_2003}. The latter was postulated in \cite{karlin_perfect_1999} on the basis of the {entropy maximum principle} and was recently derived also by coarse-graining of molecular dynamics \cite{wagner_integer_2018}. The present new derivation highlights the unique renormalizability of LBM with entropic equilibrium.

\begin{figure}[t!]
	\centering
	\includegraphics[width=0.9\linewidth,keepaspectratio]{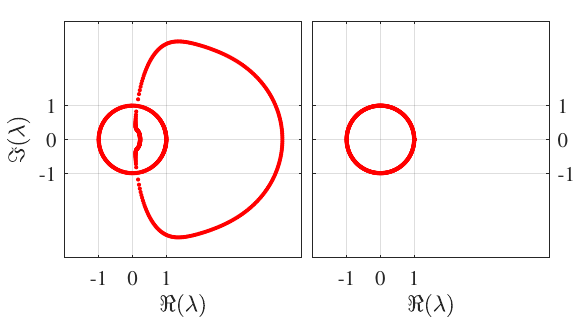}
	\caption{Root locus of the characteristic polynomial of the D$1$Q$3$ LBGK model with $\beta=0.9994$ and $u=1$. Left: Isotropic pressure; Right: Asymptotically free pressure  \eqref{eq:Pstar_entropic}.}
	\label{Fig:root_locus_poly_af_U1}
\end{figure}

{While the necessary stability condition \eqref{eq:condition1} concerns the long-wave limit, 
	the D$1$Q$3$ LBGK system allows for analytical study of both the necessary and \emph{sufficient} stability conditions at all wave numbers $k\in[0,\ 2\pi]$.
Necessary and sufficient conditions for linear stability of a discrete system are provided by the concept of Schur-stability of the characteristic polynomial \cite{marden1949geometry}. In our case, roots $\lambda$ of the third-order characteristic polynomial of the linearized LBGK \eqref{eq:LBGK} must be located within the unit disk in the complex plane, $|\lambda|\le 1$. 
To that end, we use a modified Jury table algorithm \cite{jury1988modified,choo2009schur} to identify the conditions of Schur-stability analytically, 
details can be found in the Supplementary Material \cite{supplement}.
The analysis demonstrates that  the necessary and sufficient conditions for the linear stability of the D$1$Q$3$ LBGK system are 
independent of the wave-number. This proves that the hydrodynamic limit stability condition \eqref{eq:condition1} is both necessary and sufficient for linear stability of the D$1$Q$3$ LBGK.
The effect of the choice of the pressure on the Schur-stability is illustrated in Fig.\ \ref{Fig:root_locus_poly_af_U1}: While
 $\pi^{*}=\varsigma^2$ leads to $|\lambda|\geq 1$ for some of the roots and thus to instability, the asymptotically free pressure \eqref{eq:Pstar_entropic} guarantees $|\lambda|\leq 1$ even for $|u|=1$.}

{The above necessary stability condition \eqref{eq:condition1} remains valid in higher dimensions.  
Here we consider only the asymptotically free pressure \eqref{eq:Pstar_entropic}. For the LBGK model on the D$2$Q$9$ lattice,
at the Navier--Stokes order, the non-equilibrium pressure tensor becomes,}
\begin{equation}
    \bm{\pi}^{\rm neq} = -\frac{1-\beta}{2\beta}\rho\varsigma^2\left[\left(\bm{\mathcal{A}}\odot\bm{\nabla}\bm{u}\right) + {\left(\bm{\mathcal{A}}\odot\bm{\nabla}\bm{u}\right)}^\dagger \right], \label{eq:Pi_noneq_af}
\end{equation}
where $\odot$ is the Hadamard (component-wise) product of matrices, while the matrix $\bm{\mathcal{A}}$ reads,
\begin{equation}\label{eq:visc_tens}
    \bm{\mathcal{A}} = \begin{bmatrix}
        \mathcal{A}_{xx}  && \pi^{*}_{xx}/\varsigma^2\\
        \pi^{*}_{yy}/\varsigma^2 && \mathcal{A}_{yy}
    \end{bmatrix}.
\end{equation}
{Here the off-diagonal components are defined by the pressure \eqref{eq:Pstar_entropic} as $\pi^*_{\alpha\alpha}=\pi^*(u_{\alpha})$ while the diagonal components are defined by the viscosity factor \eqref{eq:A} as $\mathcal{A}_{\alpha\alpha}=\mathcal{A}(u_{\alpha})$. Thus, with the properties of the function $\mathcal{A}$ already specified, all components of the matrix $\bm{\mathcal{A}}$ governing the decay rates of both the normal and the shear modes are non-negative in the entire range of flow velocity $|u_{x,y}|\le 1$.  Note that, the asymptotics at small velocity, $\bm{\mathcal{A}}\to\bm{{1}}-{\rm diag}\{(2/3)(u_\alpha/\varsigma)^2\}$, where $\bm{{1}}$ is the matrix with all components equal to one, is {the same} for both the isotropic and the asymptotically free pressure. Together with the linearity of the rate-of-strain $\sim\bm{\nabla}\bm{u}$, the remaining anisotropy is of the order $\sim u^3$ and is a universal consequence of the "cubic anomaly" due to the aforementioned lattice constraint. At the same time, another anomalous term of order $\sim u^3$ appears in the nonequilibrium pressure tensor due to compressibility error (proportional to $\mathcal{B}_{\alpha\alpha}=\mathcal{B}(u_{\alpha})$, cf.\ \eqref{eq:B}) when the isotropic pressure is used. For the asymptotically free pressure, the latter error is not present in \eqref{eq:Pi_noneq_af}, thus it is more accurate 
than the isotropic pressure. This is not surprising because the asymptotically free pressure was derived from the no-compressibility-error condition. Moreover, the remaining leading-order anomaly in \eqref{eq:visc_tens} can be eliminated by renormalization, similar to the D$1$Q$3$ case above, albeit within a multiple relaxation time setting rather than the LBGK. This is beyond the scope of this Letter.}

{Schur-stability analysis of the ninth-order characteristic polynomial in two dimensions becomes cumbersome, hence we probe linear stability of the LBGK by numerically solving the eigen-value problem.
Linear stability domain of the LBGK with the entropic equilibrium of Eq. \eqref{eq:Pstar_entropic} is compared to the isotropic pressure case in Fig. \ref{Fig:stability_all} for a perturbation aligned with the $x$-axis, $\bm{k}=(k_x,0)$, $k_x\in[0,2\pi]$ (see Supplemental Material \cite{supplement} for a generic wave-number perturbation). Also included in the comparison is the second-order polynomial equilibrium \cite{qian1992lattice} obtained by retaining terms of order $u_{\alpha}u_{\beta}$ in the expansion of the equilibrium populations. It is apparent that LBGK with the asymptotically free equilibrium  \eqref{eq:Pstar_entropic} is {unconditionally linearly stable}: Stability domain extends to the entire range of flow velocity $|u|\le 1$ and is {independent} of the viscosity $\nu$.  The two other LBGK with the isotropic pressure behave differently: First, stability domain is limited by the velocity $|u^{\max}|=1-1/\sqrt{3}$, the value at which isotropic pressure violates the necessary stability condition \eqref{eq:condition1}. Above this value, no amount of viscosity can stabilize the LBGK system. Second, the stability domain  shrinks to nil with the decrease of the viscosity.
All that is in marked contrast to the asymptotically free LBGK.
}
\begin{figure}[t!]
	\centering
	\includegraphics[width=6cm,keepaspectratio]{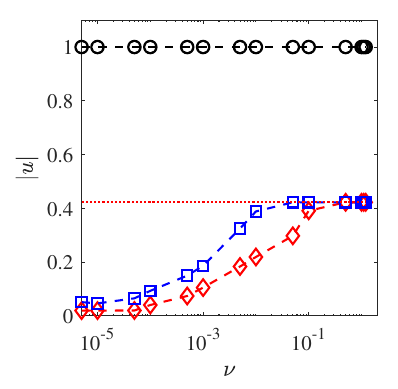}
	\caption{{Linear stability domain of D$2$Q$9$ LBGK with different equilibria. Maximal attainable flow velocity vs. viscosity \eqref{eq:viscosity}. 
	Red with diamond markers: Second-order polynomial equilibrium \cite{qian1992lattice}; Blue with square markers: Product-form equilibrium \eqref{eq:generic_isothermal_EDF} with isotropic pressure; Black with circular markers: Product-form equilibrium \eqref{eq:generic_isothermal_EDF} with asymptotically free pressure \eqref{eq:Pstar_entropic}. Horizontal dotted line: $|u^{\max}|=1-1/\sqrt{3}$.}}
	\label{Fig:stability_all}
\end{figure}

{
In summary, seminal work on quantum chromodynamics \cite{gross1973ultraviolet,politzer1973reliable} teaches us that perturbative computations at low energies in strongly coupled systems are only possible with asymptotic freedom at high energies. 
Lattice Boltzmann systems can be regarded as strongly coupled in lieu of constraints on particles' velocities imposed by the lattice. 
Thus, stable simulations at low flow velocities may require asymptotic freedom at high velocities.}

{In order to test this hypothesis, a rigorous new approach to LBM was developed in this Letter. Coupling parameters were identified as normal and shear modes while the coarse-graining brought about the necessary stability condition and which shows that, indeed, the asymptotic freedom must be guaranteed by the equilibrium. It was rigorously shown that the entropic equilibrium satisfies the asymptotic freedom and is uniquely renormalizable. Moreover, for the LBGK model, the necessary conditions are also sufficient. With the asymptotically free equilibrium, the LBGK is unconditionally linearly stable.

}

{A conventional remedy to LBGK instability invokes a concept of  multiple relaxation times (MRT), that is, the relaxation to the equilibrium proceeds at different rates for different moments of the distribution function \cite{succi_lattice_2002,kruger_lattice_2017}. However, the above analysis shows that the necessary stability condition rather concerns the equilibrium itself. Indeed, asymptotic freedom was derived within the hydrodynamic limit which is a common fixed point of any MRT. 
	We have performed numerical stability test on a variety of conventional MRT models 
	to find that all of them are bound to fail when the flow velocity exceeds the same maximum $|u^{\max}|=1-1/\sqrt{3}$ \cite{hosseini_development_2020}. None of MRT models with conventional equilibrium produces unconditionally stable LBM.}
	
The practical outcome of this Letter is a rigorous algorithm for the construction of unconditionally stable lattice Boltzmann models based on identification of modes as coupling parameters (cf.\ Eq.\ \eqref{eq:modes_D1}), analysis of conditions for their asymptotic freedom (cf.\ Eq.\ \eqref{eq:condition1}) and solving the resulting renormalization equations (cf.\ Eq.\ \eqref{eq:no_compressibility_error}). This perspective on the lattice Boltzmann construction may be particularly useful for compressible and multiphase flow LBM simulations \cite{chikatamarla_entropy_2006,mazloomi_entropic_2015}. In a broader sense, a relation between asymptotic freedom and the entropy maximum principle, which was demonstrated in the above example, may uncover new insights in statistical physics.
\begin{acknowledgments}
This work was supported by European Research Council (ERC) Advanced Grant  834763-PonD. 
Computational resources at the Swiss National  Super  Computing  Center  CSCS  were  provided  under the grant  s1222.
\end{acknowledgments}

\bibliography{references}

\end{document}